\documentclass[%
 reprint, twocolumn,
showpacs,
 amsmath,amssymb,
 aps,
 prb,
]{revtex4-1}

\usepackage{graphicx}
\usepackage{dcolumn}
\usepackage{bm}

\begin{document}

\title{Characterizing Short Necklace States in Logarithmic Transmission Spectrum of Strongly Localized Systems}
\author{Liang Chen$^{1,2}$}
\author{Xunya Jiang$^1$}
\email{xyjiang@mail.sim.ac.cn}
\affiliation{ $^1$State Key Laboratory of Functional Materials for
Informatics, Shanghai Institute of Microsystem and Information
Technology, CAS, Shanghai 200050, China} \affiliation{ $^2$Graduate
School of Chinese Academy of Sciences, Beijing 100049, People's
Republic of China}

\begin{abstract}
High transmission plateaus exist widely in the logarithmic
transmission spectra of localized systems. Their physical origins
are short chains of coupled-localized-states embedded inside the
localized system, which are dubbed as ``short necklace states''. In
this work, we define the essential quantities and then, based on
these quantities, we investigate the short necklace states'
properties \emph{statistically} and \emph{quantitatively}. Two
different approaches are utilized and the results from them agree
with each other very well. In the first approach, the typical
plateau-width and the typical order of short necklace states are
obtained from the correlation function of \emph{logarithmic}
transmission. In the second approach, we investigate statistical
distributions of the peak/plateau-width measured in
\emph{logarithmic} transmission spectra. A novel distribution is
found, which can be exactly fitted by the summation of two Gaussian
distributions. These two distributions are the results of sharp
peaks of localized states and the high plateaus of short necklace
states. The center of the second distribution also tells us the
typical plateau-width of short necklace states. With increasing the
system length, the scaling property of typical plateau-width is very
special since it almost does not decrease. The methods and the
quantities defined in this work can be widely used on Anderson
localization studies.
\end{abstract}

\maketitle

\section{Introduction} The most extraordinary phenomenon of
wave transport in random media is Anderson
localization\cite{Anderson1958}, which is shared by various systems,
such as, electronic, photonic and accoustic
systems\cite{ShengSoukoulisBook}. In Anderson localized system the
ensemble average of logarithmic transmission $\left<lnT\right>$, not
transmission $\left<T\right>$, is additive with system length
$L$\cite{Abrahams1979,Anderson1980}. Naturally, the localization
length can be defined as $\xi=-2L/\left<lnT\right>$. In strongly
localized systems($L\gg\xi$), the transmission is generally small,
but also shows large fluctuations\cite{Anderson1980}. Those extreme
large $T$ values will dominate the transmission of localized
systems. In the past years, the study of the physical origins of
those large $T$ values yielded abundant results for understanding
the transport phenomena in random systems and the Anderson
localization\cite{Azbel1983,Lee1985RMP,Pendry,Tartakovskii,Bliokh2004,
Bertolotti2005PRL,Sebbah2006PRL,Bertolotti2006PRE,Bliokh2006PRL,Ghulinyan2007PRA,
Ghulinyan2007PRL,Bliokh2008PRL,Bliokh2008RMP,Vanneste2009PRA,
LiWei2009,Zhang2009PRB,Wang2008Nano,Wang2011PRB,Wang2010PRB,ShengPingPercolation,Chen2011NJP}.

Previously, it was pointed out by  M. A. Azbel et al.\cite{Azbel1983} that
the resonant transport through those localized states lying near the system center
can contribute very large $T$ values, being of order unity.
Although such a mechanism can generate high resonant transmission peak
\cite{Azbel1983,Lee1985RMP}, its contribution to
the transmission is vanishingly small in strongly localized system because
the resonant peak is exponentially sharp ($\sim e^{-L/\xi}$)\cite{Azbel1983,Lee1985RMP,Bliokh2004}.
Laterly, Pendry\cite{Pendry} and Tartakovskii\cite{Tartakovskii} independently
predicted a kind of quasi-extended state, 
called the \emph{necklace state(NS)}. The NS is formed through the
coupling between nearly-degenerated localized states which are
evenly distributed in the system. Despite the spatial overlaps
between localized states are small, \emph{because of the degenerate
coupling}, the NS contributes very wide transmission ``mini-bands'',
which can dominate the transmission of strongly localized systems.
The NSs have been demonstrated experimentally in photonic
systems\cite{Bertolotti2005PRL,Sebbah2006PRL}. Their fundamental
characters and statistical consequences to the transmission are
widely studied\cite{Bertolotti2006PRE,Ghulinyan2007PRA,
Ghulinyan2007PRL,Bliokh2008PRL,Bliokh2008RMP}. Recently, it is
demonstrated that the NSs have evident contribution to the
short-time transport of wave
package\cite{LiWei2009,Zhang2009PRB,Wang2008Nano,Wang2011PRB} and
the dynamics of fluctuations of localized waves\cite{Wang2010PRB}.
More profoundly, the number of NS can increase dramatically as approaching
the Anderson transition point, which strongly supports a modes-coupling induced
quantum percolation scenario for the Anderson localization-delocalization
transition\cite{ShengPingPercolation,Chen2011NJP}.

Even though NSs can contribute very large transmission, their
formation has rigorous conditions, i.e., the nearly-degenerated
localized states are required to be \emph{evenly distributed} inside
the system\cite{Pendry,Ghulinyan2007PRA,Bliokh2008PRL,Bliokh2008RMP}. In
general the localized states with similar frequency are not such
well distributed in a specific random configuration. In such cases,
\emph{ideal NSs}\cite{Ghulinyan2007PRA,Ghulinyan2007PRL} crossing
the whole configuration are not formed. 
But eventually, those nearly-degenerated states which are also
\emph{spatially close to each other} couple together, forming short chains of
coupled-localized-states embedded inside the configuration. Since
those coupled-localized-states have similar properties as the NS, we
call them \emph{short necklace state}(SNS)\cite{JiangArXiv}. The SNS
also manifests as high transmission plateau which has significant
transmission contribution. In transmission spectra the SNS is hardly
to be distinguished from isolated localized states since the valley
between coupled peaks seems very low. But it can be clearly
identified from the logarithmic transmission spectra, for instance,
see the coupled-peaks in Fig.1. Because of the strong coupling
between the neighboring localized states inside the SNS, on top of
the plateau there are sharp peaks and valleys, which correspond to
un-smooth changes of the transmission phase between the
coupled-peaks\cite{Ghulinyan2007PRA}.

In contrast to the NSs, the SNSs have special
properties\cite{JiangArXiv} that (1) the SNSs exist widely in every
random configuration while the NS only occurs once in millions of
random configurations with large length($L\gg\xi$); (2) the
plateau-width of the SNSs only depends on the coupling strength
between the neighboring localized states and is insensitive to the
system length. With these properties, SNSs are superior
\emph{qualitatively} compared with the NSs.
However, how to \emph{quantitatively} characterize statistical
properties and scaling behaviors of SNSs are still unexplored topic.
Even more, some essential quantities of SNS study, such as the
``half-width" of SNS plateaus in $lnT$ spectra, need to be defined
since there is no obvious physical quantity in previous studies
can directly describe SNS.

In this paper, we study the statistical manifestation of the SNS 
using physical quantities which are defined in the
\emph{logarithmic} transmission spectra.
To characterize SNS, we find out two different approaches, whose
results can be compared with each other. The first approach is based
on the correlation functions $C_{lnT}$ (defined in the $lnT$
spectra) and $C_t$ (defined in the transmission coefficient $t$
spectra). From $C_{lnT}$, we show that there is a typical frequency
correlation range of logarithmic transmission, which is explained as
the typical SNS plateau-width. From this frequency range and
compared with the results of $C_t$, we can find the most-probable
order of SNS. The second approach is from the direct measurement of
peak/plateau-width in $lnT$ spectra. We find the statistical
distribution of peak/plateau-width is quite abnormal and
can be fitted very well by the summation of two Gaussian distribution
functions, where the primary Gaussian center gives the typical width
of resonant peaks of localized states while the secondary one gives
exactly the typical width of the SNS plateaus. Excellent agreements
are found between two approaches. The dependencies of SNS's
properties on the scaling parameter $L/\xi$ are also studied. We
find the plateau-width of SNS almost does not depend on $L/\xi$
while the peak-width of localized states decays exponentially with
$L/\xi$, indicating the SNSs have more significant transmission
contribution in longer systems. 
The essential quantities defined in this work also provide new ways
for further quantitative study on statistical properties of the
transmission of Anderson localized systems.

The rest of this paper is organized as follows. In section II, we
introduce our model and basic properties of the $lnT$ spectra. In
section III, we present our numerical results and theoretical
analysis of three correlation functions: the correlation of
transmission $C_T$, the correlation of logarithmic transmission
$C_{lnT}$ and the field correlation $C_t$. We show that the
half-width of $C_{lnT}$ gives the typical plateau-width of SNS. The
most-probable order of SNS can be obtained by comparing $C_{lnT}$
and $C_t$. In section IV, we study statistical distributions of
the peak/plateau-width (defined in the $lnT$ spectra). The
probability distributions are calculated from very large numbers
of samples and can be fitted very well by the summation of
two Gaussian distributions, where the first
Gaussian center gives the typical width of resonant peaks of
localized states and the second one gives the typical
width of the SNS plateau. In section V, we discuss the system length
dependency of the SNS. Finally, a summary of this work is given in
section VI.

\begin{figure}
\includegraphics[width=1.0 \columnwidth ]{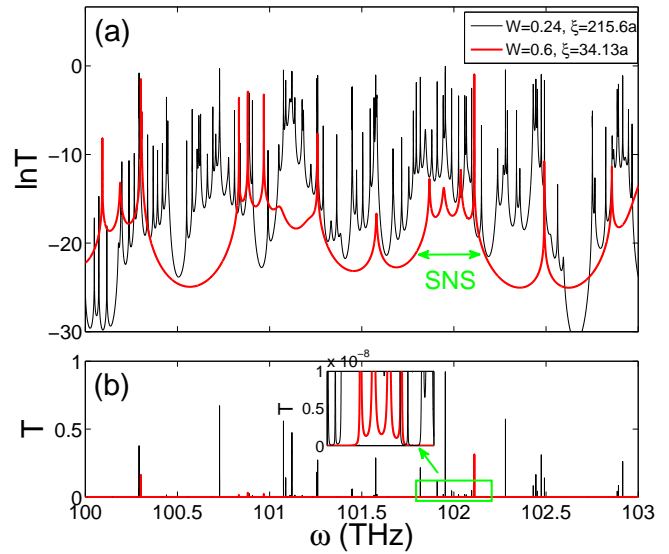}
\caption{(Color online) Typical logarithmic transmission spectra (a)
and corresponding transmission spectra (b) of $L=10\xi$ systems. Red
bold line: $\xi=34.13 a$ ($W=0.6$); black thin line: $\xi=215.6 a$
($W=0.24$). The SNS marked in (a) is enlarged in (b) for clear
observation.}
\end{figure}
\section{ The Model } We study 1D random stacks composed of
binary-dielectric layers(called A and B) with thicknesses
$d_A=d_B=500 nm$ and with refractive indices $n_A=1.0$ and
$n_B=3.0*(1+W\gamma)$, where $\gamma$ is a random number uniformly
distributed in $[-0.5,0.5]$ and $W$ gives the randomness strength.
Such a periodic on average model is an optical counterpart of the
Anderson model of electronic systems and is widely used for
localization
studies\cite{Bertolotti2006PRE,Ghulinyan2007PRA,Ghulinyan2007PRL,LiWei2009,Chen2011NJP}.
The transmission coefficients of optical waves are calculated by
standard transfer matrix method\cite{Chen2011NJP}. Without
randomness, $W=0$, the system exhibits a pass band in frequency
range $(89.13, 124.1 THz)$. Our study will focus on the range
$(100.0, 103.2 THz)$, where the localization length is almost a
constant for a certain randomness $W$($0.2 \le W \le
0.6$)\cite{Chen2011NJP}. The localization length is calculated by
$\xi=-2L/\left<lnT\right>$, where $\left<lnT\right>$ is averaged
from the $10^6$ configurations satisfying $L\gg\xi$ ($L\approx
500\xi$).

Fig.1(a) shows two typical logarithmic transmission spectra with the
same $L/\xi=10$ but with different localization lengths,
$\xi=215.6a$(black thin) and $\xi=34.14a$(red bold). We can see they
have similar average height, being approximately $-2L/\xi=-20$.
After averaged over a large number of configurations, the $lnT$
spectrum becomes a flat line exactly falls on the mean value
$\left<lnT\right>=-20$. This is a natural result according to the
single parameter scaling theorem that $lnT$ follows the Gaussian
distribution with mean value $-2L/\xi$.
An interesting phenomenon in the $lnT$ spectra is that there are
some clear plateau-structures, such as the one marked by arrows in
Fig.1(a). From the wave intensity distributions one can find those
plateaus are formed by the SNSs(i.e., from the degenerate coupling
between some spatially close localized states) embedded inside
the configuration\cite{JiangArXiv}.
Since the peaks of localized states in the spectra are extremely sharp ($\sim
e^{-L/\xi}$), the fluctuation of $lnT$ is dominated by those
plateaus. More importantly, these SNS may be essential for
understanding the Anderson transition phenomenon\cite{JiangArXiv}.
After observed a large number of $lnT$ spectra for fixed $\xi$ and
$L/\xi$, we find the plateaus appear with some intuitionistic
regularities, (1) most plateaus have similar frequency width
for a configuration; (2) the number of peaks on each plateau are
similar. For example, for most plateaus of the $\xi=34.14a$ system
shown in Fig.1, the plateau width is about $0.3\sim0.4 THz$ and
the number of peaks are about $3\sim4$. In the following we will
focus on the statistical properties of these SNSs and try to
characterize them quantitatively.

To quantitatively study the SNS, we need to find proper physical
quantities. Fig.1 shows that the plateaus of SNSs can
hardly be distinguished from localized states in a $T$ spectrum (see
Fig.1(b)), but can be clearly identified by the plateaus in the
$lnT$ spectra (Fig.1(a)). So we try to define physical quantities
in the $lnT$ spectra for the SNS study.
In the following, we will define
and study the correlation function of $lnT$ spectra in our first
approach. Then, we will define the $lnT$ peak/plateau
width--$\Gamma$ and study the statistical distribution of $\Gamma$
in our second approach. 

\section{ Correlation functions } According to the standard definition of mathematics,
we define the correlation functions of $T$ and $lnT$ in frequency
domain as:
\begin{eqnarray}
C_T(\Delta\omega)=
\frac{Cov(T_{\omega},T_{\omega+\Delta\omega})}{\sigma(T_{\omega})\sigma(T_{\omega+\Delta\omega})}
\end{eqnarray}
\begin{eqnarray}
C_{lnT}(\Delta\omega)=
\frac{Cov(lnT_{\omega},lnT_{\omega+\Delta\omega})}{\sigma(lnT_{\omega})\sigma(lnT_{\omega+\Delta\omega})}
\end{eqnarray}
where $T$ is the transmission coefficient, $\sigma$ is the standard
deviation $\sigma(x)=\sqrt{\left<(x-\left<x\right>)^2\right>}$ and
$Cov$ is the covariance
$Cov(x,y)=\left<x-\left<x\right>\right>\left<y-\left<y\right>\right>$.
Physically $C_T$ implicates the frequency correlation of transmitted
wave intensity. Its Fourier transformation corresponds to the
dynamical response at the outgoing interface, which have been
intensively studied in weakly localized
systems\cite{GenackCorrelation}. If two frequencies are on the same
transmission peak, $C_T$ is close to unity. Otherwise $C_T$ will
close to zero. Hence the half-width of $C_T$ is usually used to
characterize the linewidth of localized
states\cite{GenackCorrelation}. Similarly, on the $lnT$ spectra,
since $lnT$ of two frequencies on the same plateau are much larger
than the mean value $\left<lnT\right>$ and contribute a
significantly large $C_{lnT}$. We expect $C_{lnT}$ is close to unity
when $\Delta\omega$ is smaller than a typical plateau-width and
decays to zero when $\Delta\omega$ is larger than a typical
plateau-width.

Meanwhile, we also study the field correlation function,
\begin{eqnarray}
C_t
(\Delta\omega)=\frac {<t_{\omega} {t_{\omega+\Delta\omega}}^*
+{t_{\omega}}^*t_{\omega+\Delta\omega}>}
{<T_{\omega}>+<T_{\omega+\Delta\omega}> }
\end{eqnarray}
where $t$ is complex transmission coefficient and $T=tt^*$. The
complex amplitude of incident wave is chosen to be $E_{in}(x=0)=1$
so that at the outgoing interface $t=E(x=L)=|E|e^{i\phi}$, where $E$
is the complex electronic field and $\phi$ is its phase. Similar to
$C_T$, the physical implication of $C_t$ is simply the frequency
correlation of transmitted field, i.e., the field correlation($C_T$
is the intensity correlation). The definition of $C_t$ is the same
as that in ref.\cite{Pendry}, which could be negative valued,
depending on the averaged phase difference $\Delta\phi$ between
$\omega$ and $\omega+\Delta\omega$.
Generally, when $\Delta\omega$ crosses a resonant peak, the phase of
$t$ jumps $\pi$\cite{Pendry,Chen2011NJP}.
Hence the phase difference between $t(\omega)$ and $t(\omega+\Delta\omega)$ is
approximately $\Delta\phi=\phi(\omega)-\phi(\omega+\Delta\omega)
\approx \pi$. Then
$t_{\omega}t_{\omega+\Delta\omega}^*+t_{\omega}^*t_{\omega+\Delta\omega}=
|E_{\omega}||E_{\omega+\Delta\omega}|\cdot 2cos(\Delta\phi)$ becomes
negative and so that $C_t$ is negative. When $\Delta\omega$ gets to
the value which typically contains two peaks (plus one average distance
between two peaks), $C_t$ will be positive
since $\Delta\phi\approx 2\pi$, and so forth. Hence when increasing
$\Delta\omega$, we expect $C_t(\Delta\omega)$ to oscillate between
positive and negative. Each time $C_t(\Delta\omega)$ changes its
sign, the average number of resonant peaks in $\Delta\omega$
increases one. So $C_t(\Delta\omega)$ provides a way to check the
average resonant peak number in a certain frequency range.

\begin{figure}
\includegraphics[width=1.0 \columnwidth]{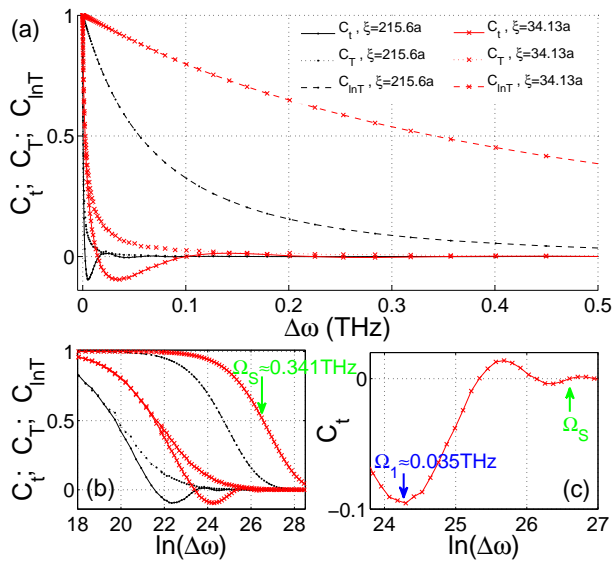}
\caption{(Color online) (a): Correlation functions of $L=10\xi$
systems calculated from $2\times 10^7$ configurations. Red:
$\xi=34.13 a$; Black: $\xi=215.6 a$. For each color, the correlation
functions from the top down are respectively $C_{lnT}$, $C_T$ and
$C_t$. (b): Same as (a) but with $\Delta\omega$ shown in logarithmic scale.
(c): Detailed view of the oscillations of $C_t$($\xi=34.13 a$). The
half-width of $C_{lnT}$, $\Omega_S$, gives the typical plateau-width
of SNS. The first minimum of $C_t$, $\Omega_1$, gives the typical
width of a resonant peak.}
\end{figure}
In our calculations we set the original frequency point $\omega=100
THz$ without lose of generality. The correlation functions
calculated from a very large number ($2\times10^7$) of
configurations for different $\xi$ are shown in Fig.2(a), where the
ratio $L/\xi$ is fixed at $10$. The solid, dashed and dotted curves
respectively represents $C_T$, $C_t$ and $C_{lnT}$. The red curves
marked by crosses correspond to the smaller $\xi$ system and the
black curves marked by solid dots correspond to the larger $\xi$
system. In Fig.2(b) the transverse axis is shown in logarithmic
scale.

Let us first discuss $C_T$. Fig.2(a) shows $C_T$ is a very singular
function at the origin $\Delta\omega\rightarrow 0$. The linewidth
$\Delta\omega$ of localized states, corresponding to
$C_T(\Delta\omega)=0.5$, is larger in the smaller $\xi$ systems.
This is in consistent with the observation in Fig.1 that the
linewidth of resonant peaks in smaller $\xi$ system is larger.
Detailed data gives that the halfwidth of the larger $\xi$ system
(black solid curve) is about $0.382GHz$ and the smaller $\xi$ system
(red solid curve) is $2.483GHz$. This is also quantitatively in
consistent with the observation from the $lnT$ spectra.

$C_{lnT}$ is a much smoother function than $C_T$ in the
$\Delta\omega\rightarrow 0$ limit. Take the $\xi=34.13 a$ system
(red curves) for example. $C_{T}$ rapidly falls to zero near
$ln(\Delta\omega) \approx 24$. $C_{lnT}$ exhibits a plateau at the
origin, then drops to $0.5$ at $ln(\Delta\omega) \approx 26.6$,
i.e.,
its half-width is about $0.341THz$, being much larger than that of
$C_T$. Such contrast reflects the different geometry properties
between the $T$ and $lnT$ spectrum. On the $T$ spectrum, $C_T$ falls
to zero typically when two frequencies are not on the same peak.
Since the peaks of localized states are exponentially sharp, $C_T$
decreases rapidly as increasing $\Delta\omega$. However, on the
$lnT$ spectrum, there exist many plateaus formed by SNSs, as shown
in Fig.1(a). Those plateaus are usually higher than the average
transmission background $\left<lnT\right>$. When two frequencies are
on the same plateau but not the same peak, it will still contribute
a significantly large value to $C_{lnT}$. Hence \emph{the halfwidth
of $C_{lnT}$, which is denoted as $\Omega_S$ in this paper,
approximately gives the typical plateau-width of SNS in $lnT$
spectrum}.

With the help of $C_t$, we can find more detailed information of
SNS, such as the average SNS order, which is the average number of
coupled localized states in one SNS. Similar to $C_T$, $C_t$ also
shows sharp singularity at the origin and falls quickly as
increasing $\Delta\omega$. Interestingly, $C_t$ shows some
oscillations around $C_t=0$ in the region where $C_T$ falls to
nearly zero, see Fig.2(c). As discussed earlier in this section,
those oscillations can be understood from the $\pi$-phase jumps of
resonant peaks of the localized states. The different order
minimal/maximal points of the $C_t$ correspond to the frequency
ranges which can accommodate certain number of resonant peaks. We
denote the $\Delta\omega$ at the first minimum of $C_t$ as
$\Omega_1$, which is the the typical frequency width of a resonant peak.
We also denote $n$th order minimal/maximal points
 $\Omega_n$ as the average frequency range which can accommodate
$n$ resonant peaks. So $C_t$ is the ruler of the number of resonant
peaks in a frequency range. With $\Omega_n$ in our mind, we can see
that the mean width of SNS plateaus, obtained by $C_{lnT}$, can
accommodate about $3\sim4$ resonant peaks, as indicated by the green
arrows in the Fig.2(c). Actually, this average number of resonant
peaks in a SNS 
agrees very well with our direct observation of many spectra.

\begin{figure}[t]
\includegraphics[width=1.0 \columnwidth ]{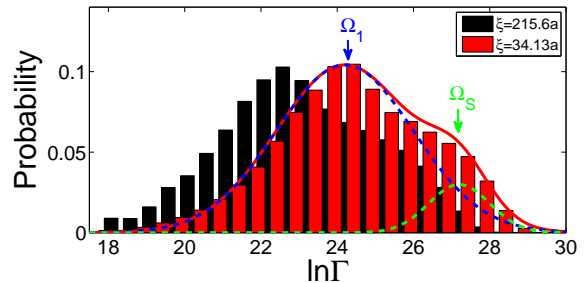}%
\caption{(Color online) Probability distribution of the
peak/plateau-width $\Gamma$(see the text for definition) measured
from logarithmic transmission spectrum. The distribution function
can be fitted very well by the summation(red solid curve) of two
Gaussian functions(dashed curves). The primary(left, blue) Gaussian
center corresponds to the typical width of resonant peaks of localized states.
The secondary(right, green) Gaussian center corresponds to the typical
plateau-width of SNS.}
\end{figure}
\section{ Peak/plateau-width statistics } To make a test and verify of
the characters of SNS obtained from $C_{lnT}$ and $C_t$, we next try
our second approach, \emph{i.e.} direct measure of the
peaks/plateaus-width in a large number of $lnT$ spectra.
Since the ensemble average of $lnT$ spectra--$\left<lnT\right>$ is
well defined, it is natural to \emph{define the peaks/plateaus-width
$\Gamma$ as the frequency interval where $lnT$ is always higher than
$\left<lnT\right>$}. From direct observation of the $lnT$ spectra,
one can find that $\left<lnT\right>$ crosses lots of sharp peaks of
localized states and fewer SNS plateaus. More precise statistical
results should be obtained from a large number of realizations.
We find that the statistical distribution of $\Gamma$ is extremely
skewed. The reason is that in Anderson localized systems the
peak-width of localized states is exponentially small and the width
of coupled-peaks also scales exponentially with the system
length\cite{Azbel1983,Pendry,Chen2011NJP}. This is very similar to
the probability distribution of the dimensionless conductance $g$,
which is extremely skewed (nearly log-normal). Early
study\cite{Anderson1980} on Anderson localization has shown that it
is better to study the additive quantity--$lng$, which is nearly
Gaussian distributed. Similarly, instead of $\Gamma$, we will study the
$ln\Gamma$ distribution, which is likely Gaussian distributed.

\begin{figure}[t]
\includegraphics[width=1.0 \columnwidth ]{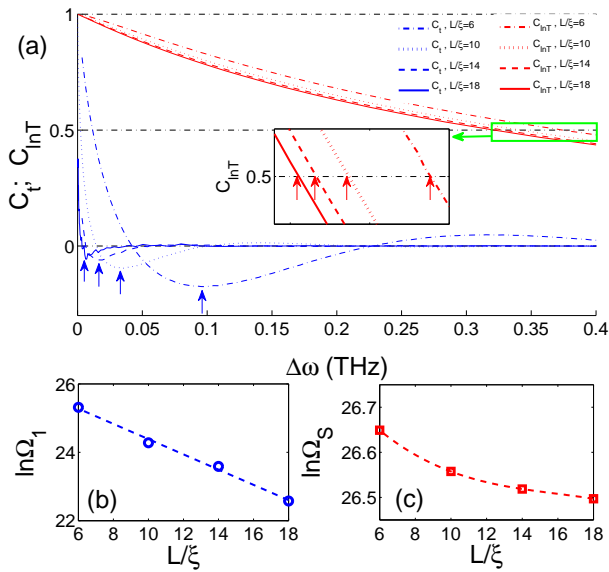}%
\caption{(Color online) (a): $C_t$ (lower four blue curves) and
$C_{lnT}$ (upper four red curves) for different $L/\xi$ systems. For
each set of four curves, $L/\xi = 6,10,14,18$ from left to right.
The inset gives detailed view at the halfwidth of $C_{lnT}$. (b):
The typical peak-width of localized states $\Omega_1$ (marked by
blue arrows in (a)) as a function of $L/\xi$.
(c): The typical plateau-width of SNS $\Omega_S$ (marked by red
arrows in (a)) as a function of $L/\xi$. $\Omega_1$ decays
exponentially as increasing $L/\xi$ while $\Omega_S$ is insensitive
to $L/\xi$.}
\end{figure}
We have measured $\Gamma$ in $10^5$ spectra with high frequency
precision that can distinguish each resonant peak in the $L=10\xi$
systems\cite{note3}. The probability distributions of $ln\Gamma$ for
two different $\xi$ systems are shown in Fig.3.
Take the $\xi=34.13a$ system as the example. (All the following
discussions also apply to the $\xi=215.6a$ system.) The probability
distribution is roughly Gaussian-like, where a clear maximum is
found at $ln\Gamma \approx 24.3$($\Gamma\approx0.035THz$). This is
exactly the $ln(\Omega_1)$ at first minimum of $C_t$ shown in Fig.2.
As shown before, this value corresponds to the typical frequency
range occupied by one single resonant peak. 
Such a result is coincident with our direct observation of the $lnT$
spectra that the most probable peaks/plateaus crossed by
$\left<lnT\right>$ are those single peaks. Suppose the spectrum
contains only sharp peaks but none plateaus. $ln(\Delta\omega)$
should be nearly Gaussian distributed with a single maximum at the
mean value. However, the real probability distribution of $ln\Gamma$
shows a strange shoulder at $ln\Gamma \approx 26.6$($\Gamma\approx
0.341THz$). Comparing with Fig.2 we find it exactly corresponds to
the halfwidth of $C_{lnT}$, $\Omega_S$, which is just the typical
width of SNS plateau, referring to the physical meaning of
$C_{lnT}$. Actually, the measured probability distribution can be
fitted very well by the summation of two Gaussian distributions, as
shown by the curves in Fig.3. The primary(left) Gaussian center gives
the width of single resonant peaks while the secondary(right) one
gives exactly the typical width of the SNS plateau. Such a strange
probability distribution implies substantial SNS plateaus with
similar frequency widths contribute significantly to the probability
distribution function of $ln\Gamma$, resulting the characteristic
shoulder of SNS.

It is first time to see directly from the statistical distribution
that the SNS is clearly distinguished from other localized states.
From the distribution, we can see that the logarithmic plateau-width
of SNS is Gaussian distributed and its mean value agrees excellently
with the results of correlation functions in third section, as
expected. So, both $C_{lnT}$ and the probability distribution of
$ln\Gamma$ provides proper physical values for characterizing the
SNS. 

\section{ System length dependencies } Both the correlation
functions and the probability distribution of $ln\Gamma$ suggest
that the SNS favors a specific frequency width and a specific
number(order) of localized states. But those studies are done with
certain $L/\xi$. Next, to drive this conclusion further, we will
calculate the correlation functions 
in the systems with different $L/\xi$. Fig.4(a) shows $C_t$ (the
lower four blue curves) and $C_{lnT}$ (the upper four red curves)
functions for several $L/\xi$ values, where $\xi$ is fixed at $34.13
a$ ($W=0.6$) and $L/\xi=6,10,14,18$ from left to right. It clearly
shows the peak width $\Omega_1$ of resonant peaks of
localized states (marked by blue arrows) decreases
dramatically as increasing $L$ while the decrease of plateau-width
$\Omega_S$(halfwidth of $C_{lnT}$ marked by red arrows) is much
smaller. $\Omega_1$ and $\Omega_S$ versus $L/\xi$ are shown in Fig.4(b) and
(c). We can see $\Omega_1$ decays exponentially with $L$ ($\sim
e^{-L/\xi})$, as expected from traditional understanding\cite{Azbel1983,Lee1985RMP,Bliokh2004}.
However, the decrease of $\Omega_S$ is almost negligible and obviously much
slower than $\Omega_1$. Actually, the plateau-width of SNS is
determined by the intrinsic coupling strength(repulsing distance)
between localized states, which almost does not depend on
$L$\cite{JiangArXiv}.
Such a distinction naturally results a picture that, as increasing
$L$ more and more, the peaks of localized states become so sharp
that they are almost undetectable, but the SNS plateaus are still
there. In our numerical statistics, the number(order) of localized
states increases nearly linearly with $L/\xi$. In
Ref.\cite{JiangArXiv} it is further argued that those SNSs (called
intrinsic short necklace states) have main contribution to the
fluctuation of transmission and also affect the value of
localization length.

\section{ Summary }
In summary, we have defined the basic quantities for SNS study and
investigate the statistical properties of SNS in the strongly
localized systems. We find two approaches to quantitatively study
SNS properties. The first approach is based on the correlation
functions and the second one is based on direct measurements of the
peak/plateau-width in logarithmic transmission spectra. In the first
approach, we defined the correlation function $C_{lnT}$ in $lnT$
spectra and show that the typical width of SNS plateaus can be
characterized by the half-width of $C_{lnT}$. And, with the help of
$C_{t}$ (correlation function of transmission coefficient $t$), the
most-probable order of SNS can be obtained. In the second approach,
we defined the peak/plateau-width $\Gamma$ in $lnT$ spectra and
studied the probability distribution of $ln{\Gamma}$ directly measured
from a large number of spectra.
The probability distribution of $ln{\Gamma}$ shows a novel shoulder
and it can be fitted very well by the summation of two Gaussian
distributions. We showed that the first one is from the distribution
of localized state peaks and the second one is from the plateaus of
SNS. The center of the second distribution gives the average
frequency width of SNS plateaus, which agrees very well with the
value obtained from correlation function. As increasing the system
length $L$, the plateau-width of SNS decays very slowly, compared
with the exponentially-decayed peak-width of localized states.
Finally, we note that the methods we used in our paper is not
limited to the transport properties of localized system. They may be
used for studying other statistical quantities which have similar
properties with the transmission of localized systems.

This work is supported by the NSFC (Grant Nos. 11004212, 11174309,
and 60938004), and the STCSM (Grant Nos. 11ZR1443800 and
11JC1414500).


\begin{thebibliography} {s1}
\bibitem{note3} Note that each spectrum covers a wide frequency range, where
a lot of peaks and plateaus will be cutted by $\left<lnT\right>$. Such that
the ensemble of $10^5$ spectra is considerably large, with about
$10^7$ sample peaks/plateaus.
\end{thebibliography}
\end{document}